\documentclass[12pt]{article}
\usepackage{amsmath,amssymb,wrapfig,tabularx}
\usepackage{xspace}
\usepackage{epsfig}
\usepackage{graphicx}  
\title{
\vspace*{-2.3cm}
\begin{flushright}\normalsize{}\end{flushright}
\vspace{1.5cm}
\Large \textbf{\sc How to preserve symmetries with cut-off\\
regularized integrals?}\vspace*{1.0cm}
\author{\large\textbf{T.~Varin~$^a$\footnote{varin@ipnl.in2p3.fr}, D.~Davesne~$^a$, M.~Oertel~$^b$, and
M.~Urban~$^c$}\\[0.5cm]
$^a$\normalsize\emph{Institut de Physique Nucl{\'e}aire de Lyon,
Universit{\'e} Claude Bernard Lyon 1,}\\
\normalsize\emph{4 rue Enrico Fermi, 69622 Villeurbanne Cedex, France}\\
$^b$\normalsize\emph{Laboratoire de l'Univers et de ses Th{\'e}ories,
Observatoire de Paris, 92195 Meudon, France}\\
$^c$\normalsize\emph{Institut de Physique Nucl{\'e}aire, CNRS and
Univ.\@ Paris-Sud 11, 91406 Orsay Cedex, France}\\}
\date{}}

\begin{document}  
\setcounter{page}{1}  
\maketitle

\begin{abstract}  
We present a prescription to calculate the quadratic and
logarithmic divergent parts of several integrals employing a
cutoff in a coherent way, i.e. in total agreement with symmetry
requirements. As examples we consider one-loop Ward identities for QED and a
phenomenological chiral model.
\end{abstract}
    
\section{\sc Introduction}  

In quantum field theory, ultraviolet (UV) divergences always emerge
from momentum integration in the loops. In order to identify the
physical content of the theory under consideration, one has to
subtract these divergences with a definite regularization procedure.
Among them, the dimensional regularization is probably the most
popular one because it is easy to use and respects all the symmetry
properties of the initial lagrangian \cite{'tHooft:1972fi,
Weinberg:1995mt, Weinberg:1996kr}. The latter aspect is of fundamental
importance and could justify by itself the choice of dimensional
regularization.

Nevertheless, the introduction of an explicit cut-off is sometimes
advantageous, for example as soon as one considers renormalization
group equations in the Wilsonian approach \cite{Weinberg:1996kr,
HY:PRep} or effective theories \cite{Georgi:1994qn}. In that context,
the problem concerning the non-respect of symmetries arises. A
three-momentum cut-off has been proved to be useful, e.g., in the case
of the Nambu-Jona-Lasinio model (see, e.g., Ref.\@ \cite{NJL3dcutoff})
but it does not respect Lorentz invariance. This can be avoided by
using a four-momentum cut-off, but the latter suffers from the fact
that it violates gauge invariance, too. This can be seen for instance
from the fact that the one-loop photon polarization tensor in QED is
no longer transverse. The problem has been solved by Schwinger within
his proper-time approach \cite{Schwinger:1948yk}. However, his
formalism is very different from the usual Feynman diagram
technique. In the literature there exists a modified version of the
proper-time regularization which can directly be applied to Feynman
diagrams \cite{Zinn-Justin}, but as shown later in this paper, the
results are ambiguous since they depend on the way the propagators are
written. On purely phenomenological grounds, the problem is sometimes
``solved'' by simply imposing gauge invariance at hand, i.e., if we
keep the example of the photon polarization tensor in QED, by
projecting it onto the transverse part. This method is of course not
very satisfactory.

Finally, another method which respects gauge invariance is that by
Pauli and Villars \cite{Pauli-Villars}. The masses of the fictitious
particles which are introduced by this method can be interpreted in
the sense of a cut-off \cite{Itzykson:1980rh}. However, as soon as the
theory allows for particles having different masses propagating in the
same loop, this method becomes impractible.  In conclusion, a
regularization scheme with an explicit cut-off dependence and
preserving symmetries would be very desirable. In this paper, we will
not give the ultimate solution of this long-standing problem. However,
within the context of renormalization group equations, it is often
sufficient to look at the divergent parts of the integrals. A
prescription to calculate these divergent parts within a cutoff
scheme preserving symmetries is thus on its own an interesting
problem. We emphasize in particular the importance of a consistent
treatment of quadratic divergences, which in the case of QED always
cancel at the end of the calculation, but which survive in other
theories and can give rise to a very strong running of the model
parameters within the renormalization group
\cite{HY:PRep,Dienes:1998vg,VarinWelzel}.

We will show through several examples how we can
conciliate the explicit dependence on a cut-off and the requirements
of symmetry. In Sect.\@ 2, we will start by giving an example where
the proper-time method fails. In Sect.\@ 3, we give a prescription
which allows to conciliate all constraints. Then, in the remaining
sections, we will give some illustrations through several examples
(QED in Sect.\@ 4, a phenomenological chiral model in Sect.\@ 5).

\section{\sc Proper-time approach}
\label{firstapproach_method}

As already mentioned in the introduction, there are essentially three
regularization methods implying a cut-off: the naive (three or four
dimensional) momentum cut-off, the method by Pauli-Villars and the
proper-time one. For the reasons detailed in the introduction we
are concentrating on a modified version of the latter, which can
immediately be applied to the calculation of Feynman
diagrams. However, as we are going to show, ambiguities may arise with
this method.

To be more explicit, let us start by giving two expressions which can
appear in a one loop calculation:
\begin{gather}
A=\int \frac{d^4k}{i(2\pi)^4}\,
\frac{k^2}{(k^2-m^2)^2}\,,
\label{int_A}
\\ 
B=\int \frac{d^4k}{i(2\pi)^4}\,
\frac{1}{(k^2-m^2)} + 
\int \frac{d^4k}{i(2\pi)^4}\,
\frac{m^2}{(k^2-m^2)^2}\,.
\label{int_B} 
\end{gather}
Although both expressions are divergent, we expect a useful
regularization procedure to guarantee the equality $A=B$. Once the
Wick rotation is done, the proper-time regularization method as stated
in the literature \cite{Zinn-Justin} consists in rewriting the
integrals and introducing a cut-off $\Lambda$ {\it via}\footnote{in
this paper, we will only consider UV divergences}:
\begin{equation}
\frac{\Gamma(n)}{(k^2\,+\,m^2)^n}
\,= 
  \int_0^\infty d\tau\,\tau^{n-1}\,e^{-\tau\,(k^2\,+\,m^2)}
  \,\longrightarrow \int_{1/\Lambda^2}^\infty d\tau\,\tau^{n-1}\,
  e^{-\tau\,(k^2\,+\,m^2)}\,,
\end{equation}
where $\Gamma$ is the usual Euler function.

In this paper we are only interested in the divergent part of the
integrals, i.e. the parts which stay finite in the limit $\Lambda
\to \infty$.  Applying the proper-time method to (\ref{int_A}) and
(\ref{int_B}), one obtains for the divergent contributions:\\
\begin{gather}
A_{div} =
  -\frac{2}{(4 \pi)^2}(\Lambda^2-m^2\,\ln\Lambda^2)\,,
\\
B_{div} =
  -\frac{1}{(4 \pi)^2}(\Lambda^2-2m^2\,\ln\Lambda^2)\,,
\end{gather}
i.e., the proper-time regularization procedure breaks the formal
equality $A=B$. The logarithmic term is the same but the coefficient
for the quadratic divergence depends on the way the integrals are
written. Of course, one might argue that rewriting (\ref{int_A}) in two
parts is not correct, since it is divergent, and decide to forbid such
a manipulation for the computation of divergences. Such a rule would
be arbitrary but acceptable if one could be sure that the results are
in agreement with all symmetry requirements. Actually, we will see in
the following sections that this is not the case.

More generally, this example exhibits the fact that some ambiguities
can arise, so that a special care has to be taken in the computation
of divergences. Therefore, we have to proceed differently in order to
obtain a consistent dependence on $\Lambda$. Such a method is
developed in the next section.

\section{\sc Consistent Approach}
\label{section_method}

When dealing with a quantum field theory containing scalar and vector fields,
we are confronted with integrals of the typical form
\begin{equation}
\int \frac{d^dk}{i(2\pi)^d}\frac{k^a}{(k^2-m^2)^b}\,.
\label{general_int}
\end{equation}
Here it is supposed that Feynman parameters have already been
introduced, such that the ``mass'' $m$ might depend on several
parameters $x_1,...,x_n$. The integrals over the Feynman parameters
are omitted since they are irrelevant for our discussion. Tensor
integrals with $k^\mu k^\nu\cdots$ in the numerator are transformed
using
\begin{gather}
\int d^d k \, k^\mu k^\nu \, f(k^2)= \frac{g^{\mu\nu}}{d}\,
  \int d^d k \, k^2 \, f(k^2)
\label{tensorintegral}
\\
\int d^d k \, k^\mu k^\nu k^\rho k^\eta \, f(k^2)=
  \frac{g^{\mu\nu}g^{\rho\eta}+g^{\mu\rho}g^{\nu\eta}
    +g^{\mu\eta}g^{\nu\rho}}{d(d+2)}\,
  \int d^d k \, k^4 \, f(k^2)
\label{tensorintegral2}
\end{gather}
and analogous rules in the case of more than four indices. In the case
of fermion fields, the same integrals appear after the traces over
gamma matrices have been evaluated. For our purposes, $b$ and $a$ can
take the values $b = 1,\dots, 4$ and $a = 0,\dots, 2b-2$, such that the
integral (\ref{general_int}) is at most quadratically
divergent. Despite we want to calculate our results in $d=4$, we leave
$d$ unevaluated for reasons which will be explained below.

The key point is not to try to regularize all these integrals
separately, but to deduce them from a single one. We introduce
\begin{equation}
I(\alpha,\beta) = \int \frac{d^dk}{i(2\pi)^d}
\frac{1}{(\alpha k^2-\beta m^2)}~,
\label{Ialphabeta1}
\end{equation}
$\alpha$ and $\beta$ being some arbitrary parameters (which will be
taken equal to $1$ at the end of the calculation). It is easy to see
that all the integrals of the form (\ref{general_int}) can be
rewritten as linear combinations of $I(\alpha, \beta)$ and its
derivatives. For instance, the expressions $A$ and $B$ defined by
Eqs.\@ (\ref{int_A}) and (\ref{int_B}) can be written as
\begin{equation}
A=\left. -\frac{\partial}{\partial \alpha}\,I(\alpha,\beta)
  \right\vert_{\alpha =\beta = 1,d=4}\,.
\label{Aderiv}
\end{equation}
and
\begin{equation}
B = \left. I(\alpha,\beta)\right\vert_{\alpha =\beta = 1,d=4} 
  + \left.\frac{\partial}{\partial \beta}I(\alpha,\beta)
    \right\vert_{\alpha =\beta = 1,d=4}
\label{Bderiv}
\end{equation}

It is straightforward to show the formal relation
\begin{equation}
I(\alpha,\beta)= \alpha^{-d/2}\int \frac{d^dk}{i(2\pi)^d}\,
\frac{1}{(k^2-\beta m^2)} = \alpha^{-d/2}\,I(1,\beta)\,.
\label{alphadependence}
\end{equation}
and the remaining integral $I(1,\beta)$ can be evaluated in $d=4$ with the
help of the proper-time method \cite{Zinn-Justin}
\begin{equation}
\left. I(\alpha = 1,\beta)\right\vert_{d=4} = -\frac{1}{(4\pi)^2}\left(
  \Lambda^2 - \beta m^2\ln\Lambda^2\right)+\dots
\end{equation}
(the dots represent terms which stay finite for $\Lambda\to\infty$),
and hence
\begin{equation}
I_{div}(\alpha,\beta) = -\frac{\alpha^{-d/2}}{(4\pi)^2}
  \left(\Lambda^2 - \beta m^2\ln\Lambda^2\right)\,.
\label{Ialphabeta2}
\end{equation}
For reasons which will become clear later, we have not yet set $d=4$
in the prefactor $\alpha^{-d/2}$. One could hope that, if one derived
all needed integrals from this single one by taking derivatives with
respect to $\alpha$ and $\beta$, everything would be
consistent. However, there still remains a subtlety as we will see
now.

From Eq.\@ (\ref{Aderiv}) we obtain
\begin{equation}
A_{div} =
  -\frac{1}{(4\pi)^2} \left(\frac{d}{2}\Lambda^2-\frac{d}{2}m^2\ln\Lambda^2
  \right)\,.
\label{Afinal}
\end{equation}
and from Eq.\@ (\ref{Bderiv})
\begin{equation}
B_{div} = -\frac{1}{(4 \pi)^2} (\Lambda^2-2 m^2 \ln\Lambda^2)\,.
\label{Bfinal}
\end{equation}
If we set $d=4$, we see that the formal equality $A=B$ is
violated. However, we can correct this by keeping $d=4$ only in the
term $\propto \ln\Lambda^2$, but setting $d=2$ in the coefficient of
the term $\propto \Lambda^2$. If we do so, the expressions for
$A_{div}$ and $B_{div}$ coincide. This prescription can be generalized
to all other integrals under consideration. Although this prescription
seems to be rather arbitrary, it allows us to obtain a consistent set
of regularized elementary integrals such that the final result is
independent of the way the non-regularized integrals are
written. These integrals are listed in the appendix.

To generalize the preceding result to loops containing derivative
couplings, let us now consider a more complicated example:
\begin{equation}
C^{\mu\nu} = \int\frac{d^dk}{i(2\pi)^d}\,
  \frac{k^2 k^\mu k^\nu}{(k^2-m^2)^3}\,.
\end{equation}
With the help of Eqs.\@ (\ref{tensorintegral}) and
(\ref{Ialphabeta1}), this integral can be written as
\begin{equation}
C^{\mu\nu} = \frac{g^{\mu\nu}}{d}
  \int\frac{d^dk}{i(2\pi)^d}\frac{k^4}{(k^2-m^2)^3}
  = \frac{g^{\mu\nu}}{2d}\,\left.\frac{\partial^2}{\partial\alpha^2}
  I(\alpha,\beta)\right|_{\alpha=\beta=1}\,.
\end{equation}
Note that we leave $d$ undetermined until the end of the
calculation. Using Eq.\@ (\ref{Ialphabeta2}), we obtain
\begin{equation}
C_{div}^{\mu\nu} = -\frac{(d+2)g^{\mu\nu}}{8(4\pi)^2}\,(\Lambda^2- 
  \beta m^2 \ln \Lambda^2)\,,
\end{equation}
and then, following the prescription given above, we finally set $d=2$
in the term $\propto \Lambda^2$ and $d=4$ in the term $\propto
\ln\Lambda^2$:
\begin{equation}
C_{div}^{\mu\nu} = -\frac{g^{\mu\nu}}{4(4\pi)^2}\,(2\Lambda^2-
  3m^2 \ln \Lambda^2)\,.
\end{equation}

This method seems very unusual concerning the way to treat the
dimensionality of the divergences. However, Veltman
\cite{Veltman:1980mj} already noticed that quadratic divergences are
associated with $d=2$ in the context of dimensional regularization,
whereas the logarithmic part has to be treated in dimension $d=4$.
More recently, Harada \textit{et al.} used a similar trick
\cite{HY:PRep,Harada:2000at, Harada:2001rf, Hidaka:2005ac} in order to
derive the renormalisation group equations for different
models of hadronic effective theories. They show \cite{HY:PRep} that
one cannot obtain correct results if one simply works in dimension
$d=4$ during the calculation.
In fact, the first integrals listed in our appendix have been obtained
in Ref.\@ \cite{HY:PRep} by calculating them in dimensional
regularization, expanding the results around $d=2$ and $d=4$, and
identifying the resulting poles according to the rules
\begin{align}
\frac{1}{2-d}\to&\frac{\Lambda^2}{8\pi}\,,\\
\frac{1}{4-d}\to&\frac{\ln\Lambda^2}{2}\,.
\end{align}
We mention that with the help of Eqs.\@ (\ref{tensorintegral}) and
(\ref{tensorintegral2}) we can reproduce in this way all the other
integrals in our appendix, too. Within this dimensional regularization
method, the remaining finite parts of the integrals can in principle
be calculated, but they are independent of $\Lambda$ and therefore
irrelevant for the renormalization group equations.

Although we are not giving a formal proof, the examples in the next
sections show that our method satisfies symmetry requirements
automatically. Let us now come to the first example, the QED
lagrangian.

\section{\sc QED vacuum polarization at one loop} 
\label{section_QED}

In order to test the procedure previously developed, it is important
to consider a simple gauge theory. This is why we start with the
regularization of the QED lagrangian in this part. With standard
notations, the lagrangian reads:
\begin{equation}
{\cal L}= \bar{\psi} (i\partial\llap/ - m) \psi 
  -\frac{1}{4}F_{\mu\nu} F^{\mu\nu} 
  - e A^\mu \bar{\psi} \gamma_\mu \psi 
  -\frac{\lambda}{2} (\partial_{\mu} A^\mu)^2\,.
\end{equation}
When one writes explicitly the gauge invariance of the generating
functional, one can obtain some general relations and constraints for
$n$-points Green functions, called the Ward identities.

One of these identities concerns the transversality of the photon
correlation function represented in Fig.\@ 1a.
\begin{figure}
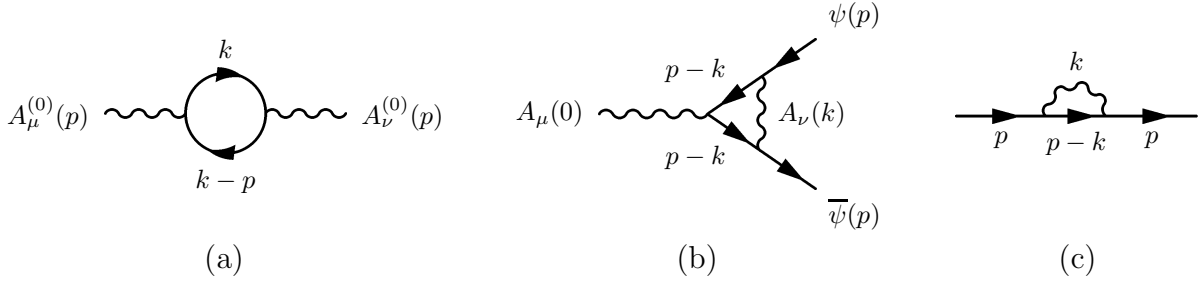

\begin{center}
\begin{tabular}{c@{\hspace{10mm}}c@{\hspace{10mm}}c}
\raisebox{5mm}{\includegraphics*[scale=1]{QED.epsi}}&
\includegraphics*[scale=1]{Ward.epsi}&
\raisebox{1cm}{\includegraphics*[scale=1]{Ward2.epsi}}\\
(a) & (b) & (c)
\end{tabular}
\end{center}
\caption{One-loop Feynman diagrams for vacuum polarization
$\Pi^{\mu\nu}(q)$, vertex correction $\Gamma^\mu(p,p)$, and electron
self-energy $\Sigma(p)$ in QED.}
\end{figure}
Using standard Feynman rules, it reads:
\begin{equation}
-i\,\Pi_{\mu \nu} (p) = -(-ie)^2 \int \frac{d^4k}{(2\pi)^4} Tr \left[
\gamma_\mu\, \frac{i}{k \hbox{\hskip -0.2 true cm}\slash -
  p \hbox{\hskip -0.2 true cm}\slash -m} \,\gamma_\nu\,
\frac{i}{k \hbox{\hskip -0.2 true cm}\slash -m} 
\right]\,.
\label{QED_SE}
\end{equation}
After introducing the Feynman parameters, we can rewrite
(\ref{QED_SE}) as follows:
\begin{equation}
\Pi_{\mu \nu} (p) = 4e^2 \int \frac{d^4k}{i(2\pi)^4} \int_0^1 dx \,
\frac{2 k_\mu k_\nu +[m^2 + x (1-x) p^2 - k^2] g_{\mu\nu}
  -2 x (1-x) p_\mu p_\nu}{(k^2-\Delta^2)^2}\,,
\end{equation}
where $\Delta^2=m^2- x (1-x) p^2$. Using the integrals listed in the
appendix, we finally obtain
\begin{equation}
\Pi_{\mu\nu} = \frac{e^2}{12 \pi^2}\, (p^2 g_{\mu \nu} - p_{\mu} p_{\nu})\,
\ln\Lambda^2 + \dots
\end{equation}
(again, the dots represent terms which stay finite for $\Lambda\to\infty$),
so that we recover immediately the transversality of the photon, as
required by the $U(1)$ gauge symmetry. A direct computation with a
naive cut-off or the proper-time method does not give this result.  It
should also be noticed that during the calculation quadratic
divergences appear, but in the final result they cancel, as it should
be.

Another quite simple and remarkable consequence of gauge invariance is
the relation between the vertex function $\Gamma_\mu(p,p)$ and the
electron self-energy $\Sigma(p)$, depicted in Figs. 1b and 1c,
respectively. The corresponding Ward identity reads:
\begin{equation}
\Gamma_\sigma (p,p) = - \frac{\partial}{\partial p ^\sigma}\,\Sigma(p)\,,
\end{equation}
where
\begin{equation}
-i \Sigma (p) = (-i e)^2 \int \frac{d^4k}{(2\pi)^4} \frac{1}{i} \left[
\frac{g_{\mu\nu}}{k^2} + (\frac{1}{\lambda}-1)\frac{k_{\mu}k_{\nu}}{k^4}
\right]
\gamma^\mu \frac{i}{p \hbox{\hskip -0.2 true cm}\slash -
k \hbox{\hskip -0.2 true cm}\slash -m} \gamma^\nu
\end{equation}
and
\begin{equation}
-i e \Gamma _\sigma(p,p) = (-i e)^3
\int\frac{d^4k}{(2\pi)^4}\frac{1}{i} \left[ \frac{g_{\mu \nu}}{k^2} +
(\frac{1}{\lambda}-1)\frac{k_{\mu}k_{\nu}}{k^4} \right]
\gamma^\mu
\frac{i}{p \hbox{\hskip -0.2 true cm}\slash -
k \hbox{\hskip -0.2 true cm}\slash -m}
\gamma_\sigma\frac{i}{p \hbox{\hskip -0.2 true cm}\slash
- k \hbox{\hskip -0.2 true cm}\slash -m} \gamma^\nu\,.
\end{equation}

After some manipulations on gamma matrices and using the formulas
listed in the appendix, one obtains
\begin{equation}
-i \Sigma (p) = \frac{i(-i e)^2}{16 \pi^2}\ln\Lambda ^2 \left(
(\frac{1}{\lambda}+3)m -
\frac{1}{\lambda} p \hbox{\hskip -0.2 true cm}\slash \right)+\dots
\end{equation}
and
\begin{equation}
-i e \Gamma _\sigma(p,p)=\frac{-(-i e)^3}{16 \pi^2 \lambda} \ln\Lambda ^2
\gamma _\sigma+\dots\,,
\end{equation}
so that the Ward identity connecting three-point and two-point Green
functions is satisfied.

There are of course other Ward identities to be checked, even at one
loop level, but the examples given above are considered as significant
tests for the consistency of our procedure. We will therefore go on to
our second example, which is a phenomenological chiral model.

\section{\sc Phenomenological chiral model}
\label{section_sigma}

As a last illustration, we consider in this section a model which
reproduces very well the phenomenology of the hadronic vector and
axial-vector correlators in the low-energy region (see
\cite{Urban:2001ru}). This model contains $\pi$, $\sigma$, $\rho$, and
$a_1$ mesons as elementary fields. Using the notations defined in
\cite{Urban:2001ru}, we write the lagrangian in the following form
\begin{align}
\cal L=\,&\frac{1}{2}\,\partial_\mu\Phi\cdot\partial^\mu\Phi
  -\frac{\mu^2}{2}\,\Phi\cdot\Phi 
  -i g\, Y_\mu \Phi\cdot\partial^\mu\Phi
  -\frac{h_1}{2}\,Y_\mu \Phi\cdot Y^\mu \Phi
  +\frac{h_2}{4}\Phi\cdot\Phi \mathrm{tr}\, Y_\mu Y^\mu
  -\frac{\lambda^2}{4}\,(\Phi\cdot\Phi)^2
\nonumber\\ 
& +c \sigma
  -\frac{1}{8}\,\mathrm{tr}\,(\partial_\mu Y_\nu-\partial_\nu Y_\mu)
    (\partial^\mu Y^\nu-\partial^\nu Y^\mu)
  +\frac{m_0^2}{4}\,\mathrm{tr}\,Y_\mu Y^\mu
  -\frac{\xi}{4}\mathrm{tr}(\partial_\mu Y^\mu)^2\, ,
\label{Lglobal}
\end{align}
where $\Phi=\begin{pmatrix}\sigma\\ \vec{\pi}\end{pmatrix}$ are the
scalar and pseudoscalar fields, and $Y_{\mu} = \vec{\rho}_\mu\cdot
\vec{T} + \vec{a}_{1\,\mu}\cdot\vec{T}^5$ are the vector and axial
vector fields. $\vec{T}$ and $\vec{T}^5$ are $O(4)$ matrices related
to the global $SU(2)_L\times SU(2)_R$ chiral symmetry, fulfilling the
commutation relations
\begin{equation}        
[T_i\,,T_j] = i \varepsilon_{ijk}\, T_k\, ,\qquad
[T_i\,,T_j^5] = i \varepsilon_{ijk}\, T_k^5\, ,\qquad
[T_i^5\,,T_j^5] = i \varepsilon_{ijk}\, T_k\, .
\label{TCommutators}
\end{equation}
Experimental information about $\rho$ and $a_1$
mesons is obtained from electromagnetic processes and $\tau$ decay
data. As a consequence, it is necessary to introduce the photon and
the $W$ fields and thus replace the ordinary derivatives by covariant
ones, following
\begin{align}
D_\mu\Phi &=\Big(\partial_\mu-i e\,A_\mu T_3
  -\frac{i e \cos\theta_C}{\sin\theta_W}\,
  (W_{1\,\mu}T^L_1+W_{2\,\mu}T^L_2)\Big)\Phi\,,
\nonumber\\
D_\mu Y_\nu&=\partial_\mu Y_\nu-i e\,A_\mu\,\lbrack T_3\,,Y_\nu\rbrack
  -\frac{i e \cos\theta_C}{\sin\theta_W}\,
  (W_{1\,\mu}\,\lbrack T^L_1\,,Y_\nu\rbrack
  +W_{2\,\mu}\,\lbrack T^L_2\,,Y_\nu\rbrack)\,.
\label{DY}
\end{align}
In order to determine the photon self-energy, one has to evaluate the
different contributions depicted in Ref.\@ \cite{Urban:2001ru} figures 10 and 11. Thanks
to the method described in the preceding section, it is possible to
show after some tedious calculations that the total $\gamma$
self-energy is simply proportional to $q^2 g_{\mu \nu} - q_{\mu}
q_{\nu}$, as it should be. Furthermore, we have explicitly checked
that this result cannot be obtained by a direct use of the proper-time
method.

Another important test of our method is the Goldstone theorem. If
there is no explicit symmetry breaking term ($c=0$ in our case), but
the global symmetry is spontaneously broken in the vacuum, then there
has to be a massless particle which corresponds to the pion in our
case. Moreover, we have to take into account the mixing between $\pi$
and $a_1$. This calculation is highly non trivial; one has to evaluate all
the diagrams depicted in Ref.\@ \cite{Urban:2001ru} (see figure 6). Writing the
self-energy for the pion in this model, one finally obtains in the
chiral limit ($c = 0$)
\begin{equation}
\Sigma_{\pi\pi}(k^2) = \frac{3 g^2}{16\pi^2\xi}\, (1-3 \xi)\, k^2\, \ln
\Lambda^2+\dots\,,
\end{equation}
i.e., the pion self-energy is simply proportional to $k^2$ and
therefore does not destroy the Goldstone character of the pion. It is
also interesting to note that, as in the case of the photon, all
quadratic divergences have canceled in the final result. The above
expression for $\Sigma_{\pi\pi}$ does not include the resummation
of the $\pi-a_1$ mixing term. When including this mixing we also recover 
the Goldstone theorem ($\Sigma_{\pi\pi}(k^2=0) = 0$), 
but the expression for the total pion
self-energy is far more complicated, such that we refrain from giving
the full expression here. Again, this result is non trivial and we
should note that, up to now, it has only been obtained in a
dimensional regularization approach (see \cite{Urban:2001ru}).

In all these examples, we have seen that the way of computing
divergences described in the third section respects symmetry
requirements.

\section{\sc Conclusions}
\label{section_conclusion}

We have seen in this paper how to handle logarithmic and quadratic
divergences in a cut-off regularization scheme in a consistent way,
i.e., in a way satisfying constraints from symmetry requirements. For
example, we have shown that our approach was preserving the
transversality of the photon polarization tensor in QED as well as the
transversality of the photon polarization tensor and the Goldstone
theorem in the case of a phenomenological chiral model including
$\pi$, $\sigma$, $\rho$, and $a_1$ mesons. In all these examples, the
quadratic divergences cancel at the end of the calculation due to the
symmetry properties of the model. Recently our method
has also been applied to QED with one extra dimension
\cite{VarinWelzel}, where the quadratic divergences survive and give
rise to a power-law in the running of the effective four-dimensional
gauge coupling. We have to stress, however, that the present method is
only designed to compute the divergent parts of the integrals. 

Our aim is now to use this regularization scheme in order to derive
renormalization group equations for the phenomenological chiral model
described in the last section. In fact, it is interesting
to incorporate the scalar degree of freedom to extend the
results of Harada \textit{et al.}
\cite{Harada:2000at,Harada:2001rf,Hidaka:2005ac} concerning the chiral
symmetry restoration. 

\section*{\sc Acknowledgments}

We gratefully acknowledge stimulating discussions with M.\@ Ericson,
G.\@ Chanfray and R.\@ Stora.

\newpage
\section*{\sc Appendix: Table of integrals}

We present in this appendix a detailed and complete list of
regularized integrals necessary for the calculation of all integrals
described in the main text and in \cite{Urban:2001ru}. As already
mentioned, only the divergent parts are given.
\begin{align}
\int \frac{d^4k}{i(2\pi)^4}\frac{1}{(k^2- m^2)} 
=&
-\frac{1}{(4\pi)^2}\left[ \Lambda^2 - m^2\ln\Lambda^2\right]+\dots \\
\int \frac{d^4k}{i(2\pi)^4}\frac{1}{(k^2- m^2)^2} 
=&
\frac{1}{(4\pi)^2}\ln\Lambda^2+\dots  \\
\int \frac{d^4k}{i(2\pi)^4}\frac{k^2}{(k^2- m^2)^2}
=&
-\frac{1}{(4\pi)^2}\left[ \Lambda^2 - 2 m^2\ln\Lambda^2 \right]+\dots \\
\int \frac{d^4k}{i(2\pi)^4}\frac{k^\mu k^\nu}{(k^2- m^2)^2}
=&
-\frac{g^{\mu\nu}}{2(4\pi)^2}\left[ \Lambda^2 - m^2\ln\Lambda^2 \right]
  +\dots \label{itensor} \\
\int \frac{d^4k}{i(2\pi)^4}\frac{k^2}{(k^2- m^2)^3}
=&
\frac{1}{(4\pi)^2}\ln\Lambda^2+\dots  \\
\int \frac{d^4k}{i(2\pi)^4}\frac{k^\mu k^\nu}{(k^2- m^2)^3}
=&
\frac{g^{\mu\nu}}{4(4\pi)^2}\ln\Lambda^2+\dots  \\
\int \frac{d^4k}{i(2\pi)^4}\frac{k^4}{(k^2- m^2)^3}
=&
-\frac{1}{(4\pi)^2}\left[ \Lambda^2 - 3 m^2\ln\Lambda^2 \right]+\dots \\
\int \frac{d^4k}{i(2\pi)^4}\frac{k^2k^\mu k^\nu}{(k^2-m^2)^3}
=&
-\frac{1}{4(4\pi)^2}\left[ 2 \Lambda^2 -3 m^2\ln\Lambda^2 \right]+\dots \\
\int \frac{d^4k}{i(2\pi)^4}\frac{k^\mu k^\nu k^\rho k^\eta}{(k^2- m^2)^3}
=&
-\frac{1}{8(4\pi)^2}\left[g^{\mu\nu}g^{\rho\eta}+
  g^{\mu\rho}g^{\nu\eta}+g^{\mu\eta}g^{\nu\rho} \right]
  \left[ \Lambda^2 - m^2\ln\Lambda^2 \right] +\dots \\
\int \frac{d^4k}{i(2\pi)^4}\frac{k^4}{(k^2- m^2)^4}
=&
\frac{1}{(4\pi)^2}\ln\Lambda^2+\dots  \\
\int \frac{d^4k}{i(2\pi)^4}\frac{k^2 k^\mu k^\nu}{(k^2- m^2)^4}
=&
\frac{g^{\mu\nu}}{4(4\pi)^2}\ln\Lambda^2+\dots  \\
\int \frac{d^4k}{i(2\pi)^4}\frac{k^\mu k^\nu k^\rho k^\eta}{(k^2- m^2)^4}
=&
\frac{1}{24(4\pi)^2}\left[g^{\mu\nu}g^{\rho\eta}+ g^{\mu\rho}g^{\nu\eta}
  +g^{\mu\eta}g^{\nu\rho} \right] \ln\Lambda^2+\dots  \\
\int \frac{d^4k}{i(2\pi)^4}\frac{k^6}{(k^2- m^2)^4}
=&
-\frac{1}{(4\pi)^2}\left[ \Lambda^2 - 4 m^2\ln\Lambda^2 \right]+\dots  \\
\int \frac{d^4k}{i(2\pi)^4}\frac{k^4 k^\mu k^\nu}{(k^2- m^2)^4}
=&
-\frac{g^{\mu\nu}}{2(4\pi)^2}\left[ \Lambda^2 - 2 m^2\ln\Lambda^2 \right]
  +\dots  \\
\int \frac{d^4k}{i(2\pi)^4}\frac{k^2 k^\mu k^\nu k^\rho k^\eta}{(k^2- m^2)^4}
=&
-\frac{1}{24(4\pi)^2}\left[g^{\mu\nu}g^{\rho\eta}+ g^{\mu\rho}g^{\nu\eta}
  +g^{\mu\eta}g^{\nu\rho}\right]
  \left[ 3 \Lambda^2 - 4 m^2\ln\Lambda^2 \right]+\dots 
\end{align}



\clearpage
\begin{thebibliography}{99}  

\bibitem{'tHooft:1972fi} 
  G.~'t Hooft and M.J.G.~Veltman, Nucl.\ Phys.\ B {\bf 44} (1972) 189.
\bibitem{Weinberg:1995mt}
  S.~Weinberg, {\it The Quantum Theory of Fields}, Vol.\@ 1:
  {\it Foundations} (Cambridge University Press, 1995).
\bibitem{Weinberg:1996kr}
  S.~Weinberg, {\it The Quantum Theory of Fields}, Vol.\@ 2:
  {\it Modern applications} (Cambridge University Press, 1996).
\bibitem{HY:PRep} M.~Harada and K.~Yamawaki, Phys.\ Rep.\ {\bf 381}, 1
  (2003).
\bibitem{Georgi:1994qn}
  H.~Georgi, Ann.\ Rev.\ Nucl.\ Part.\ Sci.\ {\bf 43} (1993) 209.
\bibitem{NJL3dcutoff} 
  S.~P.~Klevansky,
  Rev.\ Mod.\ Phys.\  {\bf 64} (1992) 649.
\bibitem{Schwinger:1948yk}
  J.~Schwinger, Phys.\ Rev.\ {\bf 74} (1948) 1439.
\bibitem{Zinn-Justin}
  J.~Zinn-Justin, {\it Quantum Field Theory and Critical Phenomena}
  (Oxford Science Publications, 1993).
\bibitem{Pauli-Villars}
  W.~Pauli and F.~Villars, Rev.\ Mod.\ Phys.\ {\bf 21} (1949) 434.
\bibitem{Itzykson:1980rh}
  C.~Itzykson and J.B.~Zuber, {\it Quantum Field Theory} (Mc Graw-Hill,
  1980).
\bibitem{Dienes:1998vg}
  K.R.~Dienes, E.~Dudas, and T.~Gherghetta, Nucl.\ Phys.\ B {\bf 537}
  (1999) 47.
\bibitem{VarinWelzel}
  T.~Varin, J.~Welzel, A.~Deandrea, and D.~Davesne, arXiv:hep-ph/0610130
  (2006).
\bibitem{Veltman:1980mj}
  M.J.G.~Veltman, Acta Phys.\ Polon.\ B {\bf 12} (1981) 437.
\bibitem{Harada:2000at}
  M.~Harada and K.~Yamawaki, Phys.\ Rev.\ D {\bf 64} (2001) 014023.
\bibitem{Harada:2001rf}
  M.~Harada and K.~Yamawaki, Phys.\ Rev.\ Lett.\ {\bf 87} (2001) 152001.
\bibitem{Hidaka:2005ac}
  Y.~Hidaka, O.~Morimatsu and M.~Ohtani, Phys.\ Rev.\ D {\bf 73} (2006)
  036004.
\bibitem{Urban:2001ru}
  M.~Urban, M.~Buballa and J.~Wambach, Nucl.\ Phys.\ A {\bf 697} (2002) 338.
\end{thebibliography}
\end{document}